\documentclass[reqno,a4paper,final]{amsart}
\usepackage{amssymb,amstext,amsthm,eucal}
\usepackage{bbold}
\usepackage{array}
\usepackage[latin1]{inputenc}
\usepackage[notref,notcite]{showkeys}
\usepackage{hyperref}
\hypersetup{%
  pdftitle   = {Parallelisable heterotic backgrounds},
  pdfkeywords = {heterotic string, parallelism, supersymmetry, Lie
    groups, plane waves},
  pdfauthor  = {Jos\'e Figueroa-O'Farrill, Teruhiko Kawano, Satoshi
    Yamaguchi},
  pdfcreator = {\LaTeX\ with package \flqq hyperref\frqq}
}
\newcommand{\half}{\tfrac{1}{2}}
\renewcommand{\d}{\partial}
\newcommand{\fg}{\mathfrak{g}}
\newcommand{\fh}{\mathfrak{h}}
\newcommand{\fm}{\mathfrak{m}}
\newcommand{\fso}{\mathfrak{so}}
\newcommand{\fspin}{\mathfrak{spin}}
\newcommand{\SO}{\mathrm{SO}}
\newcommand{\Spin}{\mathrm{Spin}}
\newcommand{\SL}{\mathrm{SL}}
\newcommand{\SU}{\mathrm{SU}}
\newcommand{\RR}{\mathbb{R}}
\newcommand{\ZZ}{\mathbb{Z}}

\newcommand{\eS}{\mathcal{S}}
\newcommand{\be}{\boldsymbol{e}}
\newcommand{\bx}{\boldsymbol{x}}
\newcommand{\btheta}{\boldsymbol{\theta}}
\newcommand{\blambda}{\boldsymbol{\lambda}}
\newcommand{\1}{\mathbb{1}}
\newcommand{\Fo}{\overset{\circ}{F}}
\newcommand{\nablo}{\overset{\circ}{\nabla}}
\newcommand{\Cl}{\mathrm{C}\ell}
\DeclareMathOperator{\ad}{ad}
\DeclareMathOperator{\CW}{CW}
\DeclareMathOperator{\AdS}{AdS}
\DeclareMathOperator{\dvol}{dvol}
\DeclareMathOperator{\Tr}{Tr}
\DeclareMathOperator{\tr}{tr}
\allowdisplaybreaks[1]
\begin{document}
\title{Parallelisable heterotic backgrounds}
\author[Figueroa-O'Farrill]{Jos\'e Figueroa-O'Farrill}
\address{School of Mathematics, University of Edinburgh, Scotland,
  United Kingdom}
\email{j.m.figueroa@ed.ac.uk}
\author[Kawano]{Teruhiko Kawano}
\author[Yamaguchi]{Satoshi Yamaguchi}
\address{Department of Physics, University of Tokyo, Hongo, Tokyo
  113-0033, Japan}
\email{kawano@hep-th.phys.s.u-tokyo.ac.jp}
\email{yamaguch@hep-th.phys.s.u-tokyo.ac.jp}
\thanks{EMPG-03-14, UT-03-26, \texttt{hep-th/0308141}}
\begin{abstract}
  We classify the simply-connected supersymmetric parallelisable
  backgrounds of heterotic supergravity.  They are all given by
  parallelised Lie groups admitting a bi-invariant lorentzian metric.
  We find examples preserving 4, 8, 10, 12, 14 and 16 of the 16
  supersymmetries.
\end{abstract}
\maketitle
\tableofcontents

\section{Introduction}
\label{sec:intro}

In this note we present a classification of simply-connected
supersymmetric parallelisable backgrounds of heterotic string theory;
equivalently, supersymmetric parallelisable backgrounds up to local
isometry.  We work in the supergravity limit and construct all
parallelisable backgrounds of ten-dimensional type I supergravity
coupled to supersymmetric Yang--Mills.  Parallelisable backgrounds
have been studied recently in the context of type II string theory
\cite{SSJ,JMFPara,KYPara}.  The interest in parallelisable backgrounds
stems from the fact that string theory on such backgrounds should be
exactly solvable.  This is clear for type II string theory and also
for the heterotic backgrounds without gauge fields, since as we will
show the dilaton is linear and hence can be described by a Liouville
theory \cite{NS5Brane}, whereas the geometry is that of a parallelised
Lie group and hence can be described by a WZW model \cite{WW}.  For
the backgrounds with gauge fields and hence a nonlinear dilaton, the
nonlinearity is only a function of the null coordinate $x^-$ and, as
in the homogeneous plane waves of \cite{PRTwaves,BOLhpw} this also
should be exactly solvable.

This paper is organised as follows.  In Section~\ref{sec:hetsugra} we
briefly set out the problem by defining the theory under
consideration: the supergravity limit of heterotic string theory.
Under the assumption of parallelisability we write down the equations
of motion for bosonic backgrounds and the conditions for preservation
of some supersymmetry.  In Section~\ref{sec:parallelisms} we briefly
review the parallelised geometries which will be the focus of this
paper.  In Section~\ref{sec:susy} we derive some useful consequences
of preservation of supersymmetry which underlie the bulk of the
analysis.  In particular we show that any supersymmetric background
with trivial gauge fields must have a constant or linear dilaton.
This suggests breaking the problem into two, depending on whether or
not we turn on the gauge fields.  In Section~\ref{sec:linear} we
classify the supersymmetric backgrounds with a linear (or constant)
dilaton.  This follows closely the analysis in \cite{JMFPara,KYPara}
for type II backgrounds.  In Section~\ref{sec:gauge} we classify the
supersymmetric backgrounds with nonlinear dilaton, hence with
nontrivial gauge fields.  Our results are displayed in a number of
tables, particularly Table~\ref{tab:summary} which summarises the
results.  We conclude in Section~\ref{sec:moduli} with some comments
on the moduli space of parallelisable backgrounds.  Finally,
Appendix~\ref{sec:clifford}, includes our conventions for Clifford
algebras and the Clifford action of differential forms on spinors.

\section{Heterotic supergravity}
\label{sec:hetsugra}

The field theory limit of the heterotic string \cite{HeteroticString}
is given by ten-dimensional $N{=}1$ supergravity \cite{ChamseddineN=1}
coupled to $N{=}1$ supersymmetric Yang--Mills \cite{superYM}.  This
theory was constructed in \cite{ChaplineManton} generalising the
construction in \cite{10dsugraMaxwell} for the abelian theory.  The
bosonic field content consists of a ten-dimensional lorentzian metric
$g$, a real scalar $\phi$ (the dilaton), a $3$-form $H$ and a gauge
field $A$ whose field-strength $F$ can be thought of as the curvature
two-form of a connection on a gauge bundle $P$ with gauge group $E_8
\times E_8$ or $\Spin(32)/\ZZ_2$.

In the string frame, the bosonic action on a spacetime $M$ is given by
\begin{equation}
  \label{eq:action}
  I = \int_M \dvol_g e^{-2\phi} \left( R + 4 |d\phi|^2 - \half
    |H|^2 - \tfrac{N}{2} |F|^2 \right)~,
\end{equation}
where the norms $|-|^2$ are the natural (indefinite) norms induced
from the metric:
\begin{equation*}
  |d\phi|^2 = g^{\mu\nu} \d_\mu \phi \d_\nu \phi\qquad
  |H|^2 = \tfrac16 H_{\mu\nu\rho} H^{\mu\nu\rho}\qquad
  |F|^2 = \half \Tr F_{\mu\nu} F^{\mu\nu}~,
\end{equation*}
and where $\Tr$ is an invariant metric on the Lie algebra $\fg$ of the
gauge group.

The $3$-form $H$ is not necessarily closed, and instead one has
\begin{equation*}
  dH = \tfrac{N}{2} \Tr F \wedge F - 2\alpha' \tr \Omega \wedge
  \Omega~,
\end{equation*}
where $\Omega$ is the curvature of the (unique) metric connection with
torsion $3$-form $H$.  We will be specialising to parallelisable
backgrounds; that is, those with $\Omega = 0$.  For these backgrounds,
this equation simplifies to
\begin{equation}
  \label{eq:dH}
  dH = \tfrac{N}{2} \Tr F \wedge F~.
\end{equation}
For parallelisable backgrounds, the bosonic equations of motion
also simplify, and one is left with
\begin{align}
  H_{\mu\nu\rho}\nabla^\rho\phi &= 0\label{eq:Hdphi}\\
  \nabla_\mu \nabla_\nu \phi &= \tfrac{N}4 \Tr F_{\mu\rho}
  F_\nu{}^\rho\label{eq:Hessphi}\\
  4 |d\phi|^2 &= |H|^2 + \tfrac{3N}{2} |F|^2\label{eq:norms}\\
  \nabla_\nu \left(e^{-2\phi} F^{\nu\mu}\right) + \left[A_\nu, e^{-2\phi}
    F^{\nu\mu}\right] &= \half H^{\mu\nu\rho} e^{-2\phi}
  F_{\nu\rho}\label{eq:div}
\end{align}
This last equation is the Yang--Mills equation in this geometry and
simply says that the gauge covariant divergence of $e^{-2\phi}F$
vanishes.

The fermionic fields in the theory consist of a gravitino $\psi$, a
dilatino $\lambda$ and a gaugino $\chi$.  Let $S_\pm$ denote spinor
bundles on $M$ associated to the real sixteen-dimensional chiral
spinor representations of $\Spin(1,9)$.  Then $\psi$ is a section of
$T^*M \otimes S_+$, $\chi$ is a $\fg$-valued section of $S_+$
(strictly speaking a section of $\ad P \otimes S_+$, where $\ad P$ is
the adjoint bundle of $P$) and $\lambda$ is a section of $S_-$.  In a
bosonic background, the supersymmetric variations of these fields are
given by
\begin{equation}
  \label{eq:susyvars}
  \begin{aligned}[m]
    \delta \psi &= \hat \nabla \varepsilon\\
    \delta \lambda &= (d\phi + \half H) \varepsilon\\
    \delta \chi &= \tfrac1{2\sqrt{2}} F \varepsilon~,
  \end{aligned}
\end{equation}
where $\varepsilon$ is a section of $S_+$, $\hat\nabla$ is the spin
connection associated to the metric connection with torsion $3$-form
$H$, and as usual, differential forms act on spinors via the Clifford
action as described in Appendix~\ref{sec:clifford} which also contains
our conventions concerning Clifford algebras and their
representations.

For the parallelisable backgrounds which are the focus of this paper,
$\hat\nabla$ is flat; hence there are no local obstructions to finding
parallel sections.  Since we will be interested only in local
metrics---equivalently, in simply-connected spacetimes---there will
not be any global obstructions either, whence we will not dwell on the
first of the above three equations except to remark the following
fact.  Since a Killing spinor is parallel with respect to a flat
connection $\hat \nabla$, it is uniquely determined by its value at
any given point.  In particular, a Killing spinor is
nowhere-vanishing.  We will use this fact implicitly in deriving
relations arising from the existence of Killing spinors.

In this paper we shall be concerned with simply-connected
parallelisable supersymmetric heterotic backgrounds $(M,g,\phi,H,F)$,
where $(M,g,H)$ is parallelisable and where $(g,\phi,H,F)$ satisfy the
equations \eqref{eq:dH}, \eqref{eq:Hdphi}, \eqref{eq:Hessphi},
\eqref{eq:norms} and \eqref{eq:div}, and such that there exists at
least one nonzero spinor $\varepsilon$ for which the variations
\eqref{eq:susyvars} vanish.

\section{Parallelisable geometries}
\label{sec:parallelisms}

In this section we will briefly review the basic facts concerning
parallelisable geometries.  For a recent treatment see
\cite{JMFPara}.  We will say that a pseudo-riemannian manifold $(M,g)$
is \emph{parallelisable} if it admits a flat metric connection with
torsion.

It is possible to list all the simply-connected parallelisable
lorentzian manifolds in any dimension.  This uses the following three
basic theorems.  The first theorem, due to Élie Cartan and Schouten
\cite{CartanSchouten1,CartanSchouten2} and to Wolf \cite{Wolf1,Wolf2},
says that the irreducible simply-connected parallelisable riemannian
manifolds are the following: the real line with the standard
metric and vanishing torsion, a simply-connected compact simple Lie
group with a bi-invariant metric and the parallelising torsion of
Cartan--Schouten, or $S^7$ with the canonical round metric and the
torsion coming from octonionic multiplication.  The second theorem due
to Wolf \cite{Wolf1,Wolf2} and Cahen and Parker \cite{CahenParker}
states that the indecomposable parallelisable lorentzian manifolds are
precisely the Lie groups with bi-invariant lorentzian metric and
parallelising torsion.  The third result is the classification of
simply-connected Lie groups admitting bi-invariant lorentzian metrics,
which follows from the structure theorem of Medina and Revoy
\cite{MedinaRevoy} (see also \cite{FSalgebra}) on indecomposable Lie
algebras admitting invariant lorentzian metrics.  The simply connected
lorentzian Lie groups are given by $(\RR,-dt^2)$, the universal cover
of $\SL(2,\RR)$ (i.e., $\AdS_3$) and a subclass of the Cahen--Wallach
\cite{CahenWallach} spaces $\CW_{2n}(\blambda)$.  For completeness we
recall their definition.

Let $M=\RR^2 \times \RR^{2n-2}$ with coordinates $(x^+,x^-,\bx)$ and
let $\left<-,-\right>$ denote the euclidean inner product on
$\RR^{2n-2}$.  Let $J:\RR^{2n-2} \to \RR^{2n-2}$ be an invertible
skew-symmetric linear map and let $\omega$ be the corresponding
$2$-form.  Let $\lambda_1 \geq \lambda_2 \geq \dots \geq
\lambda_{n-1} > 0$ denote the skew-eigenvalues of $J$.  The Lie group
$\CW_{2n}(\blambda)$ is diffeomorphic to $M$ with metric
\begin{equation}
  \label{eq:CWmetric}
  g = 2 dx^+ dx^- - \left<J\bx,J\bx\right> (dx^-)^2 +
  \left<d\bx,d\bx\right>~,
\end{equation}
and parallelising torsion given by
\begin{equation}
  \label{eq:CWtorsion}
  H = dx^- \wedge \omega~.
\end{equation}

In summary, we can now list the ingredients out of which we can build
all ten-dimensional parallelisable lorentzian geometries.  For each
one we also list properties concerning the dilaton $\phi$ and the
torsion $3$-form $H$.  These results are summarised in
Table~\ref{tab:ingredients}, whose last column follows from equation
\eqref{eq:Hdphi}.

\begin{table}[h!]
  \centering
  \setlength{\extrarowheight}{3pt}
  \renewcommand{\arraystretch}{1.3}
  \begin{small}
    \begin{tabular}{|>{$}l<{$}|>{$}l<{$}|>{$}l<{$}|}\hline
      \multicolumn{1}{|c|}{Space} & \multicolumn{1}{c|}{Torsion} &
      \multicolumn{1}{c|}{Dilaton} \\
      \hline\hline
      \AdS_3 & dH=0 \quad |H|^2 < 0 & \text{constant}\\
      \RR^{1,n},~n\geq 0 & H=0 & \text{unconstrained}\\
      \RR^n,~n \geq 1 & H = 0 & \text{unconstrained}\\
      S^3 & dH=0 \quad |H|^2 > 0 &  \text{constant}\\
      S^7 & dH\neq 0 \quad |H|^2 > 0 & \text{constant}\\
      \SU(3) & dH= 0 \quad |H|^2 > 0 & \text{constant}\\
      \CW_{2n}(\blambda) & dH=0 \quad |H|^2 = 0 & \phi(x^-)\\ \hline
    \end{tabular}
  \end{small}
  \vspace{8pt}
  \caption{Elementary simply-connected parallalisable geometries}
  \label{tab:ingredients}
\end{table}

Indeed, in the case of a Lie group, that is, when $dH=0$, equation
\eqref{eq:Hdphi} says that $d\phi$ must be central, when thought of as
an element in the Lie algebra. Since $\AdS_3$, $S^3$ and $\SU(3)$ are
simple, their Lie algebras have no centre, whence $d\phi=0$.  In the
case of an abelian group there are no conditions, and in the case of
$\CW_{2n}(\blambda)$, the Lie algebra has a one-dimensional centre
corresponding to $\d_+$, whose dual one-form is $dx^-$.  This means
that $d\phi$ must be proportional to $dx^-$, whence $\phi$ can only
depend on $x^-$.  Finally for $S^7$, equation \eqref{eq:Hdphi} says
that $\star H \wedge d\phi = 0$, which implies that $d\phi=0$.  To see
this, notice that the parallelised $S^7$ possesses a nearly parallel
$G_2$ structure and the differential forms decompose into irreducible
types under $G_2$.  For example, the one-forms corresponding to the
irreducible seven-dimensional irreducible representation $\fm$ of
$G_2$ coming from the embedding $G_2 \subset \SO(7)$, whereas the
two-forms decompose into $\fg_2 \oplus \fm$, where $\fg_2$ is the
adjoint representation which is irreducible since $G_2$ is simple.
Now, $H$ and $\star H$ both are $G_2$-invariant and hence the map
$\Omega^1(S^7) \to \Omega^2(S^7)$ defined by $\theta \mapsto \star
(\star H \wedge\theta)$ is $G_2$-equivariant.  Since it is not
identically zero, it must be an isomorphism onto its image.  Hence if
$\star H \wedge d\phi = 0$, then also in this case $d\phi = 0$.

It is now a simple matter to put these ingredients together to make up
all possible ten-dimensional combinations with lorentzian signature.
Doing so, we arrive at Table~\ref{tab:geometries} (see also
\cite{JMFPara}, where the entry corresponding to $\RR^{1,0} \times S^3
\times S^3 \times S^3$ had been omitted inadvertently and where the
entries with $S^7$ had also been omitted due to the fact that in type
II string theory $dH=0$).

\begin{table}[h!]
  \centering
  \setlength{\extrarowheight}{3pt}
  \renewcommand{\arraystretch}{1.3}
  \begin{small}
    \begin{tabular}{|>{$}l<{$}|>{$}l<{$}|}\hline
      \multicolumn{1}{|c|}{Spacetime} & \multicolumn{1}{c|}{Spacetime}\\
      \hline\hline
      \AdS_3 \times S^7 & \AdS_3 \times S^3 \times S^3 \times \RR\\
      \AdS_3 \times S^3 \times \RR^4 & \AdS_3 \times \RR^7\\
      \RR^{1,0} \times S^3 \times S^3 \times S^3 & \RR^{1,1} \times \SU(3)\\
      \RR^{1,2} \times S^7 & \RR^{1,3} \times S^3 \times S^3\\
      \RR^{1,6} \times S^3 & \RR^{1,9}\\
      \CW_{10}(\blambda) & \CW_8(\blambda) \times \RR^2\\
      \CW_6(\blambda) \times S^3 \times \RR & \CW_6(\blambda) \times \RR^4\\
      \CW_4(\blambda) \times S^3 \times S^3 & \CW_4(\blambda) \times
      S^3 \times \RR^3\\
      \CW_4(\blambda) \times \RR^6 & \\\hline
    \end{tabular}
  \end{small}
  \vspace{8pt}
  \caption{Ten-dimensional simply-connected parallelisable spacetimes}
  \label{tab:geometries}
\end{table}

\section{Some consequences of supersymmetry}
\label{sec:susy}

In this section we will derive some consequences of the existence of
nonzero Killing spinors.

The gaugino variation says that
\begin{equation*}
  F \varepsilon = 0~.
\end{equation*}
Clifford multiplying this equation by $F$ and tracing over $\fg$, we
find
\begin{equation}
  \label{eq:Fsquared}
  \Tr (F \wedge F) \varepsilon = |F|^2 \varepsilon~.
\end{equation}

At the same time, $F$ is a $\fg$-valued $2$-form.  At a fixed but
arbitrary point in $M$, $F$ defines an element in $\fso(1,9) \otimes
\fg$.  Moreover, because $F$ annihilates a nonzero spinor, it actually
defines an element in $\fh \otimes \fg$, where $\fh \subset \fso(1,9)$
is the isotropy algebra of the spinor.  Now the orbit structure of the
chiral spinor representation under $\Spin(1,9)$ is very simple: all
nonzero spinors belong to the same orbit \cite{Bryant-ricciflat}.  The
isotropy of a nonzero spinor is therefore conjugate in $\Spin(1,9)$ to
a fixed $\Spin(7) \ltimes \RR^8$ subgroup, which is described in
detail, for example, in \cite{JMWaves}.  In other words, there exists
a coframe $\btheta = (\theta^-, \theta^+, \theta^i)$ relative to which
the metric is written as
\begin{equation*}
  g = 2 \theta^+ \theta^- + \sum_i (\theta^i)^2~,
\end{equation*}
and $F$ is written as
\begin{equation*}
  F = \Fo_{-i} \theta^- \wedge \theta^i + \half \Fo_{ij} \theta^i
  \wedge \theta^j~,
\end{equation*}
where the $\Fo_{ij}$ belong to a $\fspin(7)$ subalgebra of $\fso(8)$.

The norm $|F|^2$ is independent on the coframe, so we compute it
relative to $\btheta$ and obtain
\begin{equation}
  \label{eq:normF}
  |F|^2 = \Tr \sum_{i<j} (\Fo_{ij})^2 \geq 0~,
\end{equation}
which is positive semidefinite.  Similarly, rewriting equation
\eqref{eq:Hessphi} in this coframe, we find
\begin{equation*}
  \nablo_a \nablo_b \phi = \tfrac{N}4 \Tr \sum_i \Fo_{ai} \Fo_{bi}~,
\end{equation*}
which is also positive-definite; whence in this coframe, and hence in
all, the Hessian of $\phi$ vanishes if and only if $F=0$.  In other
words, supersymmetric backgrounds with $F=0$ are precisely those with
a linear dilaton.

Finally let us remark that if $dH=0$, whence $\Tr(F\wedge F) = 0$,
equation \eqref{eq:Fsquared} implies that $|F|^2 = 0$, which by
\eqref{eq:normF} implies that $\Fo_{ij} = 0$.  In other words,
relative to the coframe $\btheta$ the only nonzero components of $F$
are $\Fo_{-i}$, so that
\begin{equation}
  \label{eq:newframe}
  F_{\mu\nu} = \left( \theta_\mu{}^{-} \theta_\nu{}^i - \theta_\mu{}^i
    \theta_\nu{}^{-} \right) \Fo_{-i}~.
\end{equation}

\section{Linear dilaton backgrounds}
\label{sec:linear}

In this section we consider (supersymmetric) backgrounds with $F=0$.
As we saw above, these are precisely the backgrounds where the Hessian
of the dilaton vanishes; whence the dilaton is (at most) linear.

First of all we notice that $S^7$ cannot appear because equation
\eqref{eq:dH} implies that $dH=0$.  Therefore the allowed backgrounds
follow \emph{mutatis mutandis} from the analysis of
\cite{JMFPara,KYPara}.  We only have to remember when counting
supersymmetries that we are dealing with $N{=}1$ supergravity.  In
practice this simply means halving the supersymmetries present in type
IIB supergravity.  We start by listing the possible backgrounds and
then counting the amount of supersymmetry that each preserves.  The
results are summarised in Table~\ref{tab:linear} and
Table~\ref{tab:summary}.

\subsection{Possible backgrounds}

\subsubsection{$\AdS_3 \times S^3 \times S^3 \times \RR$}

Here $d\phi$ can only have nonzero components along the flat
direction, which is spacelike, whence $|d\phi|^2 \geq 0$.  Equation
\eqref{eq:norms} says that $|H|^2 \geq 0$, so that if we call $R_0$,
$R_1$ and $R_2$ the radii of curvature of $\AdS_3$ and of the two
$3$-spheres, respectively, then
\begin{equation*}
  \frac{1}{R_1^2}  +  \frac{1}{R_2^2} \geq \frac{1}{R_0^2}~.
\end{equation*}
This bound is saturated if and only if the dilaton is constant.

\subsubsection{$\AdS_3 \times S^3 \times \RR^4$}

This is the limit $R_2 \to \infty$ of the above case.

\subsubsection{$\AdS_3 \times \RR^7$}

This would be the limit $R_1 \to \infty$ of the above case, but then
the inequality $R_0^{-2} \leq 0$ cannot be satisfied.  Hence this
geometry is not a background (with or without supersymmetry).

\subsubsection{$\RR^{1,9}$}

In this case $H=0$, so $|d\phi|^2 =0$.  So we can take a linear
dilaton along a null direction: $\phi = a + b x^-$, for some constants
$a,b$ say.

\subsubsection{$\RR^{1,0} \times S^3 \times S^3 \times S^3$}

The dilaton can only depend on the flat coordinate, which is timelike,
so $|d\phi|^2 \leq 0$.  However $|H|^2 > 0$, whence this geometry
is never a background (with or without supersymmetry).

\subsubsection{$\RR^{1,1} \times \SU(3)$}

Here $|H|^2 > 0$, and $d\phi$ can have components along
$\RR^{1,1}$.  Letting $(x^0,x^1)$ be flat coordinates for $\RR^{1,1}$,
we can take $\phi = a + \half |H| x^1$, for some constant $a$,
without loss of generality.

\subsubsection{$\RR^{1,3} \times S^3 \times S^3$}

Here $|H|^2 > 0$ and $d\phi$ can have components along
$\RR^{1,3}$ \cite{Khuri}.  With $(x^0,x^1,x^2,x^3)$ being flat
coordinates for $\RR^{1,3}$, we take $\phi = a + \half |H| x^1$, for
some constant $a$.

\subsubsection{$\RR^{1,6} \times S^3$}

This is the limit $R_2 \to \infty$ of the above case, where $R_2$ is
the radius of curvature of one of the spheres
\cite{DuffLu5Brane,NS5Brane}.

\subsubsection{$\CW_{2n}(\blambda) \times \RR^{10-2n}$, $n=2,3,4,5$}

In these cases $|H|^2 = 0$ and hence $|d\phi|^2 = 0$, so that it
cannot have components along the flat directions (if any).  This means
$\phi = a + b x^-$, for constants $a,b$.

\subsubsection{$\CW_{4}(\blambda) \times S^3 \times S^3$}

Here $|d\phi|^2 = 0$, whereas $|H|^2 >0$, hence there are no
backgrounds with this geometry.

\subsubsection{$\CW_{2n}(\blambda) \times S^3 \times \RR^{7-2n}$,
  $n=2,3$}

Here $|H|^2 > 0$, whence $|d\phi|^2 > 0$.  This means that we can
take $\phi = a + b x^- + \half |H| y$, where $y$ is any flat
coordinate in $\RR^{7-2n}$ and $a,b$ are constants.

\begin{table}[h!]
  \centering
  \setlength{\extrarowheight}{3pt}
  \renewcommand{\arraystretch}{1.3}
  \begin{small}
    \begin{tabular}{|>{$}l<{$}|>{$}l<{$}|}\hline
      \multicolumn{1}{|c|}{Geometry} &
      \multicolumn{1}{c|}{Dilaton} \\
      \hline\hline
      \AdS_3 \times S^3 \times S^3 \times \RR & \phi = a + \half |H| y\\
      \AdS_3 \times S^3 \times \RR^4 & \phi = a + \half |H| y\\
      \RR^{1,1} \times \SU(3) &  \phi = a + \half |H| y\\
      \RR^{1,3} \times S^3 \times S^3 & \phi = a + \half |H| y\\
      \RR^{1,6} \times S^3 & \phi = a + \half |H| y\\
      \RR^{1,9} & \phi = a + b x^-\\
      \CW_{10}(\blambda) & \phi = a + b x^-\\
      \CW_8(\blambda) \times \RR^2 & \phi = a + b x^-\\
      \CW_6(\blambda) \times S^3 \times \RR & \phi = a + b x^- + \half
      |H| y\\
      \CW_6(\blambda) \times \RR^4 & \phi = a + b x^-\\
      \CW_4(\blambda) \times S^3 \times \RR^3 & \phi = a + b x^- +
      \half |H| y\\
      \CW_4(\blambda) \times \RR^6 & \phi = a + b x^-\\
      \hline
    \end{tabular}
  \end{small}
  \vspace{8pt}
  \caption{Parallelisable backgrounds with a linear dilaton.  The
    notation is such that $y$ is a spacelike flat coordinate.}
  \label{tab:linear}
\end{table}

\subsection{Supersymmetry}

We must distinguish between three cases: $d\phi = 0$, $d\phi\neq 0$
but $|d\phi|^2 = 0$ and $|d\phi|^2 > 0$.  The results are
summarised in Table~\ref{tab:summary}.

\subsubsection{$d\phi=0$}

This can be read off from \cite{JMFPara} and we will not repeat the
analysis here.  We simply read off the results for type IIB and halve
the number of supersymmetries.

\subsubsection{$d\phi\neq 0$ and $|d\phi|^2 = 0$}

This follows from \cite{KYPara}.  Notice that this only occurs with
$\CW_{2n}(\blambda) \times \RR^{10-2n}$ for $n=2,3,4,5$ and for
$\RR^{1,9}$.  In the former cases, the dilatino variation implies an
equation of the form
\begin{equation*}
  (b + \half \omega) \cdot dx^- \varepsilon = 0~,
\end{equation*}
where the $\cdot$ means Clifford multiplication as defined in
Appendix~\ref{sec:clifford}.  Clearly half the supersymmetries will be
killed by $dx^-$, whereas the operator $b + \half \omega$ is
invertible since $b$ is real and $\omega$ is invertible and has no
real eigenvalues.  In the latter case, the dilaton variation is simply
\begin{equation*}
  dx^- \varepsilon = 0~,
\end{equation*}
whence the background is also half-BPS.

\subsubsection{$|d\phi|^2 > 0$}
\label{sec:projector}

This also follows from \cite{KYPara}.  Multiply the dilatino variation
by $d\phi$ and using that $d\phi \cdot H = - H \cdot d\phi$, which
follows from equation \eqref{eq:Hdphi}, to obtain
\begin{equation*}
  (-|d\phi|^2 \1 + \half d\phi \cdot H) \varepsilon = 0
\end{equation*}
which is equivalent to
\begin{equation*}
  \left(\half \1 + \tfrac14 \frac{H \cdot d\phi}{|d\phi|^2} \right)
  \varepsilon = 0~.
\end{equation*}
Define endomorphisms
\begin{equation*}
  \Pi_\pm = \half \1 \pm \tfrac14 \frac{H \cdot d\phi}{|d\phi|^2}~.
\end{equation*}
One sees immediately that $\Pi_+ + \Pi_- = \1$ and from equation
\eqref{eq:norms} that $\Pi_+ \cdot \Pi_- =0$, whence they are
idempotent: $\Pi_\pm^2 = \Pi_\pm$.  In other words, they are
complementary projectors and their kernels are therefore
half-dimensional.  Therefore these backgrounds are also half-BPS.

\section{Turning on the gauge fields}
\label{sec:gauge}

We now relax the condition that $F=0$ and consider backgrounds with
non-linear dilatons.  We will go through each geometry in turn and
determine which ones can carry a nontrivial gauge field.  For those
backgrounds we will then determine the amount of supersymmetry that is
preserved.  The results are summarised in Table~\ref{tab:summary}.

\subsection{Possible backgrounds}

\subsubsection{$\AdS_3 \times S^7$}

In this case the dilaton is forced to be constant, whence $F=0$, which
contradicts $dH \neq 0$.

\subsubsection{$\AdS_3 \times X$, $X\neq S^7$}

Chose coordinates $x^\mu = (x^0,x^m)$ where $x^0$ is timelike and
$x^m$ are spacelike. The timelike component of the gradient of the
dilaton must vanish: $\nabla_0 \phi = 0$.  Equation \eqref{eq:Hessphi}
then says that
\begin{equation*}
  \Tr F_{0m} F_0{}^m = 0 \implies F_{0m}=0~.
\end{equation*}
Since $dH=0$, $|F|^2 = 0$, which implies that $F_{mn}=0$.  Therefore
$F=0$ and the dilaton has to be linear.

\subsubsection{$\RR^{1,0} \times S^3 \times S^3 \times S^3$}

Here $|d\phi|^2 \leq 0$, whereas $|H|^2 > 0$ and $|F|^2 \geq
0$.  Therefore equation \eqref{eq:norms} cannot be satisfied with or
without supersymmetry.

\subsubsection{$\RR^{1,1} \times \SU(3)$}

Let us choose coordinates $(x^+,x^-)$ for $\RR^{1,1}$ and $x^\alpha$
for $\SU(3)$.  It follows from \eqref{eq:Hdphi} that $\nabla_\alpha
\phi = 0$.  Because $dH=0$, equation \eqref{eq:newframe} applies:
\begin{equation*}
  F_{\mu\nu} = \left( \theta_\mu{}^{-} \theta_\nu{}^i - \theta_\mu{}^i
    \theta_\nu{}^{-} \right) \Fo_{-i}~,
\end{equation*}
whence by \eqref{eq:Hessphi},
\begin{equation*}
  \nabla_\mu \nabla_\nu \phi = \tfrac{N}{4} \theta_\mu{}^- \theta_\nu{}^-
  \Tr \sum_i (\Fo_{-i})^2~.
\end{equation*}
Since $\nabla_\alpha \phi =0$, it follows that $\nabla_\alpha
\nabla_\alpha \phi = 0$, whence
\begin{equation*}
  (\theta_\alpha{}^-)^2 \Tr \sum_i (\Fo_{-i})^2 = 0~.
\end{equation*}
Since $F\neq 0$, we must have $\theta_\alpha{}^- = 0$, whence
$F_{\alpha\beta} = 0$; and since $|F|^2=0$, this yields the identity
\begin{equation}
  \label{eq:ip}
 \Tr F_{+-}^2 = 2 \Tr F_{+\alpha} F_-{}^\alpha~.
\end{equation}

Now since $\btheta$ is an orthonormal coframe, $\theta_\mu{}^-
\theta_\nu{}^- g^{\mu\nu} = 0$.  Since $\theta_\alpha{}^-=0$, this
implies that $\theta_+{}^- \theta_-{}^- = 0$, whence either
$\theta_+{}^- = 0$ and hence $F_{+\alpha} = 0$ or else 
$\theta_-{}^- = 0$ and hence $F_{-\alpha} = 0$.  In either case,
equation \eqref{eq:ip} implies that $F_{+-} = 0$.

Without loss of generality, let us assume that $\theta_+{}^-=0$, so
that the only surviving components of $F$ are $F_{-\alpha}$.  This
implies that the only surviving component of the Hessian of $\phi$ is
\begin{equation*}
  \nabla_- \nabla_- \phi = \tfrac{N}{4} \Tr F_{-\alpha}
  F_-{}^\alpha~.
\end{equation*}

Finally we use that equation \eqref{eq:norms} and the fact that
$|H|^2$ is constant imply that so is $\nabla_+ \phi \nabla_- \phi$.
Differentiating with respect to $\nabla_-$, we obtain
\begin{equation*}
  \nabla_+\phi \nabla_-\nabla_- \phi = 0~,
\end{equation*}
whence either $\nabla_+\phi = 0$, in which case $|d\phi|^2 = 0$
contradicting that $|H|^2>0$ and hence our assumption that $F\neq
0$, or else the Hessian of $\phi$ vanishes, whence $F=0$.  In either
case, we conclude that this geometry does not admit nontrivial gauge
fields.

\subsubsection{$\CW_4(\blambda) \times S^3 \times S^3 $}

Here $|d\phi|^2 = 0$, whereas $|H|^2 >0$ and $|F|^2 \geq 0$, so
that equation \eqref{eq:norms} cannot be satisfied.  This argument is
independent on supersymmetry, whence this is not a parallelisable
heterotic background with or without supersymmetry.

\subsubsection{$\CW_{2n}(\blambda) \times S^3 \times \RR^{7-2n}$,
  $n=2,3$}

These cases can be analysed simultaneously.  We choose standard
coordinates $(x^+,x^-,x^m)$ for $\CW_{2n}(\blambda)$, $x^\alpha$ for
$S^3$ and flat coordinates $x^I$ for $\RR^{7-2n}$.

The only components of $\nabla\phi$ which are allowed to be nonzero
are $\nabla_-\phi$ and $\nabla_I\phi$, where $\nabla_I\phi$ cannot all
vanish because $|d\phi|^2 >0$ since $|H|^2 > 0$.

Since $dH=0$, equation \eqref{eq:newframe} holds and substituting it
into equation \eqref{eq:Hessphi} we obtain
\begin{equation*}
  \nabla_\mu \nabla_\nu \phi = \tfrac{N}{4}
  \theta_\mu{}^-\theta_\nu{}^- \Tr \sum_i (\Fo_{-i})^2~.
\end{equation*}
Since we are interested in $F\neq 0$, $\Tr \sum_i (\Fo_{-i})^2 >0$.
This implies, together with $\nabla_+\phi = \nabla_m\phi =
\nabla_\alpha \phi = 0$, that $\theta_+{}^- = \theta_\alpha{}^- =
\theta_m{}^- = 0$.  Since $\btheta$ is orthonormal, $\theta_\mu{}^-
\theta_\nu{}^- g^{\mu\nu} = 0$, whence $\theta_I{}^- = 0$, whence
the only nonzero entry is $\theta_-{}^-$, which into
\eqref{eq:newframe} says that the only nonzero components of $F$
are $F_{-\mu}$.  Moreover, since $|F|^2=0$, it follows in addition
that $F_{-+}=0$.

This allows us to choose the gauge $A = A_- dx^-$, where $A_-$ does
not depend on $x^+$.  It also means that the only nonzero component of
the Hessian of $\phi$ is
\begin{equation*}
  \nabla_-\nabla_- \phi = \tfrac{N}{4} g^{ij} \Tr F_{-i} F_{-j}~,
\end{equation*}
where the indices $i,j$ label collectively all the transverse
coordinates $(x^m,x^\alpha,x^I)$.  In particular, we see that
$\nabla_-\nabla_I \phi = \nabla_I \nabla_J \phi = 0$.  From this we
see that
\begin{equation*}
  \phi = \varphi(x^-) + Q_I x^I~,
\end{equation*}
where $Q_I$ are constants obeying
\begin{equation*}
  \sum_I (Q_I)^2 = \tfrac14 |H|^2
\end{equation*}
so that equation \eqref{eq:norms} is satisfied; and $\varphi(x^-)$
satisfies the second order ODE:
\begin{equation}
  \label{eq:odebis}
  \varphi'' = \tfrac{N}{4} g^{ij} \Tr \d_i A_-\d_j A_-~,
\end{equation}
which has a unique solution for specified initial conditions at least
locally.

It remains to satisfy equation \eqref{eq:div}, which in this case
simplifies to
\begin{equation}
  \label{eq:lap}
  g^{\nu\rho} \nabla_\nu \left( e^{-2\phi} F_{\rho\mu} \right) = 0
  \implies g^{ij} \nabla_i \d_j A_- = 2 \sum_I Q_I \d_I A_-~.
\end{equation}

We now apply the transverse laplacian $\Delta^\perp = g^{ij} \nabla_i
\nabla_j$ to \eqref{eq:odebis} to obtain
\begin{equation}
  \label{eq:lapbis}
  |\nabla \d A_-|^2 + R^{ij} \Tr (\d_i A_- \d_j A_-)
  + g^{ij} \Tr \nabla_i \Delta^\perp A_- \nabla_j A_- = 0~,
\end{equation}
where $\nabla\d A_-$ stands for the transverse Hessian $\nabla_i\d_j
A_-$ of $A_-$ and $R^{ij}$ is the Ricci tensor.  After using
\eqref{eq:lap}, this equation becomes
\begin{equation*}
  |\nabla \d A_-|^2 + R^{ij} \Tr (\d_i A_- \d_j A_-)
  = - \sum_I Q_I \d_I  g^{ij} \Tr \d_i A_- \d_j A_-~.  
\end{equation*}
Now the right-hand side of this equation is zero because $g^{ij} \Tr
\d_i A_- \d_j A_-$ is proportional to $\varphi''$ which only depends on
$x^-$, whence we are left with
\begin{equation*}
    |\nabla \d A_-|^2 + R^{ij} \Tr (\d_i A_- \d_j A_-) = 0~.
\end{equation*}
Now the Ricci tensor $R^{ij}$ is only nonzero on the sphere, on which
it is positive-definite, hence the above expression becomes
\begin{equation*}
    |\nabla \d A_-|^2 + R^{\alpha\beta} \Tr (\d_\alpha A_- \d_\alpha
    A_-) = 0~,
\end{equation*}
which is manifestly positive-definite and hence will vanish if and only if
\begin{equation*}
  \nabla_i \d_j A_- = 0 \qquad\text{and}\qquad \d_\alpha A_- = 0~,
\end{equation*}
with the consequence that
\begin{equation*}
  A_- = a(x^-) + f_I(x^-) x^I~.
\end{equation*}
Finally, in order to satisfy \eqref{eq:lap}, we must demand $\sum_I
Q_I f_I(x^-) = 0$.

\subsubsection{$\CW_{2n}(\blambda) \times \RR^{10-2n}$, $n=2,3,4,5$}

These cases can also be analysed simultaneously.

Here $dH=0$, whence $|F|^2=0$ and also $|H|^2=0$, whence
$|d\phi|^2=0$.  This means that $\phi$ can only depend on $x^-$.
Therefore $\nabla_+\phi = \nabla_i \phi = 0$.  We now insert these
into equation \eqref{eq:Hessphi} to obtain the following implications:
\begin{align*}
  \nabla_+\nabla_+ \phi = 0 \implies \sum_i \Tr (F_{+i})^2 = 0 &
  \implies F_{+i} = 0\\
  \nabla_i\nabla_i \phi = 0 \implies \sum_j \Tr (F_{ij})^2 = 0 &
  \implies F_{ij} = 0\\
  \nabla_-\nabla_+ \phi = 0 \implies \Tr (F_{+-})^2 = 0 & \implies
  F_{+-} = 0~.
\end{align*}
As a result the only nonzero component of $F$ is $F_{-i}$ and this
means that we can choose a gauge where the only nonzero component of $A$ is
$A_-$.  Moreover, because $F_{+-}=0$, $\d_+ A_- = 0$.  The
Yang--Mills equation \eqref{eq:div} becomes
\begin{equation}
  \label{eq:harmonic}
  \sum_i \d_i^2 A_- = 0~,
\end{equation}
which says that $A_-$ is harmonic in the transverse coordinates.
Applying the transverse laplacian to the only nonzero component of the
Hessian of $\phi$
\begin{equation*}
  \nabla_-\nabla_- \phi = \tfrac{N}{4} \Tr \sum_i (\d_i A_-)^2
\end{equation*}
we obtain
\begin{equation*}
  \Tr \sum_{i,j} \left( \d_i A_- \d_j^2 \d_i A_- + \d_i \d_j A_- \d_i
    \d_j A_- \right) = 0~.
\end{equation*}
Using equation \eqref{eq:harmonic}, we obtain
\begin{equation*}
  \Tr \sum_{i,j} \left( \d_i \d_j A_-\right)^2 = 0~,
\end{equation*}
whence $\d_i \d_j A_- = 0$.  In other words,
\begin{equation*}
  A_-(x^-,x^i) = a(x^-) + f_i(x^-) x^i~,
\end{equation*}
where $a$ and $f_i$ are arbitrary smooth functions of $x^-$.  Finally
the dilaton $\phi(x^-)$ is obtained by solving the second order ODE
\begin{equation*}
  \phi'' = \tfrac{N}{4} \Tr \sum_i (f_i)^2~,
\end{equation*}
which, since the right-hand side is differentiable, has a unique
differentiable solution for specified initial conditions at least
locally.

\subsubsection{$\RR^{1,2} \times S^7$}

Choose coordinates $(x^-,x^+,y)$ for $\RR^{1,2}$ and $x^\alpha$ for
$S^7$.  As discussed in Section~\ref{sec:susy}, there is a coframe
$\btheta$ relative to which the only nonzero components of $F$ are
$\Fo_{-i}$ and $\Fo_{ij}$, the latter in linear combinations belonging
to a $\fspin(7)$ subalgebra of $\fso(8)$.  Moreover, since $dH\neq 0$
one has that $|F|^2 >0$ and hence the $\Fo_{ij}$ cannot all vanish.

The Hessian of the dilaton is given by
\begin{equation*}
  \nabla_\mu \nabla_\nu \phi = \tfrac{N}{4} \theta_\mu{}^a \theta_\nu{}^b \Tr
  \sum_i \Fo_{ai} \Fo_{bi}~,
\end{equation*}
where, as discussed in Section~\ref{sec:susy}, the above trace defines
a positive-definite quadratic form.  Since the dilaton obeys
$\nabla_\alpha \phi =0$, we see that $\nabla_\alpha \nabla_\alpha \phi
= 0$, whence $\theta_\alpha{}^a \Fo_{ai} = 0$ for all $\alpha,i$.
This expands to
\begin{equation}
  \label{eq:zerovec}
   \theta_\alpha{}^- \Fo_{-i} + \theta_\alpha{}^j \Fo_{ji} = 0~.
\end{equation}
Expanding $F$ relative to this frame, one finds
\begin{equation*}
  F_{\mu\nu} = (\theta_\mu{}^- \theta_\nu{}^i - \theta_\nu{}^-
  \theta_\mu{}^i) \Fo_{-i} + \theta_\mu{}^i \theta_\nu{}^j \Fo_{ij}~;
\end{equation*}
whence
\begin{equation*}
  F_{\alpha \nu} = - \theta_\nu{}^- \theta_\alpha{}^i \Fo_{-i}
\end{equation*}
after using equation \eqref{eq:zerovec}.
In particular, since $F_{\alpha\alpha} =0$, we see that either
$\theta_\alpha{}^- = 0$ or $\theta_\alpha{}^i \Fo_{-i} = 0$.
Similarly, we have that
\begin{equation*}
  F_{\alpha\beta} = \theta_\alpha{}^- \theta_\beta{}^i \Fo_{-i} = - 
  \theta_\alpha{}^i \theta_\beta{}^- \Fo_{-i}~,
\end{equation*}
whence either if $\theta_\alpha{}^- = 0$ or $\theta_\alpha{}^i
\Fo_{-i} = 0$ we see that $F_{\alpha\beta} = 0$.  In other words, this
means that $\Tr F\wedge F$ cannot have a component in $\Omega^4(S^7)$
contradicting equation \eqref{eq:dH} and the fact that here $dH$ is a
nonzero $4$-form on $S^7$.

\subsubsection{$\RR^{1,9}$}

Here we can take the coframe $\btheta$ to be a coordinate coframe:
$(dx^-, dx^+, dx^i)$ and hence $F=\Fo$, which only has components
$F_{-i}$.  This means that the Killing spinor, being annihilated by
$F$, obeys $dx^- \varepsilon = 0$.  The dilatino equation $d\phi
\varepsilon =0$ implies that $d\phi$ is proportional to $dx^-$, whence
$\phi$ only depends on $x^-$.

Since $F_{-i}$ are the only nonzero components, we can choose the
gauge $A = A_-(x^-,x^i) dx^-$, whence equation \eqref{eq:Hessphi}
becomes
\begin{equation}
  \label{eq:phidd}
  \phi'' = \tfrac{N}{4} \Tr \sum_i (\d_i A_-)^2~.
\end{equation}
The Yang--Mills equation \eqref{eq:div} is equivalent to
\begin{equation*}
  \sum_i \d_i^2 A_-=0~,
\end{equation*}
which says that $A_-$ is harmonic in the transverse space.  Applying
the transverse laplacian $\sum_i \d_i^2$ to equation \eqref{eq:phidd}
and using that $A_-$ is harmonic, we obtain
\begin{equation*}
  \Tr \sum_{i,j} (\d_i\d_j A_-)^2 = 0~,
\end{equation*}
which says that the Hessian of $A_-$ vanishes; whence $A_-$ is at most
linear in the transverse variables:
\begin{equation*}
  A_- = a(x^-) + f_i(x^-) x^i~,
\end{equation*}
where $a,f_i$ are arbitrary smooth functions of $x^-$.  Finally we
obtain the dilaton by solving the second order ODE
\begin{equation*}
  \phi'' = \tfrac{N}{4} \Tr \sum_i (f_i)^2~,
\end{equation*}
which has (at least locally) a unique solution for prescribed initial
data.

\subsubsection{$\RR^{1,3} \times S^3 \times S^3$, $\RR^{1,6} \times S^3$} 

These two cases can be analysed simultaneously.  The geometries are
both of the form $\RR^{1,n} \times X$, where $X$ is a compact
semisimple Lie group.  We will choose coordinates $x^\sigma =
(x^-,x^+,x^s)$ for the flat factor and $x^\alpha$ for the semisimple
Lie group.

Since $dH=0$ and hence $|F|^2 =0$, we know that there is a coframe
$\btheta$ relative to which equation \eqref{eq:newframe} holds and
hence that the Hessian of the dilaton is given by
\begin{equation*}
  \nabla_\mu \nabla_\nu \phi = \tfrac{N}{4} \theta_\mu{}^-
  \theta_\nu{}^- \Tr \sum_i (\Fo_{-i})^2~.
\end{equation*}
We would like to consider backgrounds with $F\neq 0$, whence the above
trace is positive.  Since $\nabla_\alpha\phi = 0$, we find that
$\theta_\alpha{}^- = 0$.  This means that the Killing spinor under
consideration is annihilated by Clifford multiplication with
$\theta^- = \theta_\sigma{}^- dx^\sigma$ which only has components in
the flat factor.

As we will now argue, this means that we can choose a coordinate
coframe $(dx^-,dx^+,dx^s)$ for the flat factor relative to which the
Killing spinor is annihilated by $dx^-$.

We first observe that Killing spinors in $\RR^{1,n} \times X$ are
parallel with respect to a product connection (with torsion).
Standard facts about holonomy representations can then be invoked to
show that parallel spinors are linear combinations of tensor products
of parallel spinors in each of the factors (see, for example,
\cite{Leistner}).  Strictly speaking this is true only when we
complexify, since otherwise the spinor representations are not
generally tensor products.

Therefore a (complex) Killing spinor in $\RR^{1,n} \times  X$ can be
written as a linear combination
\begin{equation*}
  \varepsilon =  \sum_i  \alpha_i \otimes \eta_i  
\end{equation*}
where each of the $\alpha_i$ are parallel in $\RR^{1,n}$ and the
$\eta_i$ are parallel in $X$; and where we can assume, without loss of
generality, that the $\eta_i$ are linearly independent.  The
connection defining parallel transport in $\RR^{1,n}$ is the (flat)
Levi--Cività connection; whence the $\alpha_i$ are actually constant
relative to a coordinate coframe made out of flat coordinates.

Similarly, there is a frame for the Lie group in which the $\eta_i$
are constant.  This can be seen as follows.  For a Lie group with a
bi-invariant metric there are two canonical flat connections:
corresponding to the (metric compatible) absolute parallelisms
obtained by left and right multiplication.  The corresponding
connection one-forms are the right- and left-invariant Maurer--Cartan
one-forms, respectively.  Let us choose the left-invariant
Maurer-Cartan form, for definiteness.  Then the corresponding parallel
spinor equation is
\begin{equation*}
  d \eta_i(g) + R(g^{-1} dg) \eta_i(g) = 0~,
\end{equation*}
where $R$ is the relevant spinorial representation of the Lie
algebra and we are letting $g = g(x^\alpha)$ label the points in the
group.  With some abuse of notation we can let $R$ also stand for
the spinorial representation of the Lie group and hence $R(g^{-1}dg) =
R(g)^{-1} d R(g)$, whence the above equation can be rewritten as
\begin{equation*}
  R(g)^{-1} d (R(g) \eta_i(g)) = 0~,
\end{equation*}
whose solution is manifestly given by $\eta_i(g) = R(g)^{-1}
\eta_i(e)$, with $e$ the identity element in the Lie group.

In summary, we have that
\begin{equation*}
  \varepsilon(x) =  (\1 \otimes R(g)^{-1}) \sum_i  \alpha_i
  \otimes \eta_i(e) = (\1 \otimes R(g)^{-1}) \varepsilon_0~,
\end{equation*}
where $\varepsilon_0$ is a constant spinor.  In other words, there
exists a frame and hence a dual coframe, relative to which the Killing
spinors are actually constant.  Using a constant Lorentz
transformation $\Lambda$ we can take this coframe to $\btheta =
(\theta^-,\theta^+,\theta^i)$ relative to which $\theta^-
\varepsilon_0 = 0$ and hence that the only possibly nonzero components
of $F$ are $\Fo_{-i}$.

A priori, the Lorentz transformation $\Lambda$ may not respect the
decomposition $\RR^{1,n} \times X$; that is, it may not belong to the
subgroup $\SO(1,n) \times \SO(9-n) \subset \SO(1,9)$ of the
ten-dimensional Lorentz group; however we have seen above that
$\theta^-$ is a one-form on $\RR^{1,n}$, whence $\Lambda$ can be
chosen to belong to this subgroup.  Since constant Lorentz
transformations preserve flat coordinates, we can choose $\theta^- =
dx^-$.  In other words, we have that $\theta_\mu{}^- = 0$ except for
$\theta_-{}^- = 1$.  This means that the only nonzero components of
$F$ are those of the form $F_{-s}$ and $F_{-\alpha}$.  The Hessian
equation for the dilaton says the only nonzero element of the Hessian
is $\nabla_-\nabla_-\phi$.

At this moment the rest of the analysis follows \emph{mutatis
  mutandis} the case of $\CW_{2n}(\blambda) \times S^3 \times
\RR^{7-2n}$ discussed above.

\subsection{Supersymmetry}

We must distinguish between two separate classes of supersymmetric
backgrounds with nontrivial gauge fields.

\subsubsection{Backgrounds with $|H|^2>0$}  These are the
backgrounds with an $S^3$ factor: $\RR^{1,3} \times S^3 \times S^3$,
$\RR^{1,6} \times S^3$, $\CW_6(\blambda) \times S^3 \times \RR$ and
$\CW_4(\blambda) \times S^3 \times \RR^3$.

The gaugino variation says that
\begin{equation*}
 F \varepsilon = f_I dx^I \wedge dx^- \varepsilon = f_i dx^I \cdot
 dx^- \varepsilon= 0~,
\end{equation*}
which is equivalent, for nonzero $F$, to $dx^- \varepsilon = 0$.

On a spinor $\varepsilon$ annihilated by $dx^-$, the dilatino
variation says that
\begin{equation}
  \label{eq:dilatinobis}
  (Q_I dx^I + \half H_S) \varepsilon = 0~,
\end{equation}
where $H_S$ is the three-form for the sphere(s).  The same reasoning
as in Section~\ref{sec:projector} allows us to conclude that solutions
of \eqref{eq:dilatinobis} coincide with the kernel of the projector
\begin{equation*}
  \Pi = \half \1 + \tfrac14 \frac{H_S\cdot Q}{|Q|^2}~,
\end{equation*}
where $Q = Q_I dx^I$ has norm $|Q|^2 = \tfrac14 |H_S|^2$, which is
therefore nonzero.  In summary, Killing spinors are in one-to-one
correspondence with the subspace of the chiral spinor representation
of $\Spin(1,9)$ consisting of spinors which are annihilated by $dx^-$
and by the above projector $\Pi$.  This is clearly a $4$-dimensional
subspace, whence these backgrounds are $\frac14$-BPS.

\subsubsection{Backgrounds with $|H|^2=0$}

These are the backgrounds without $S^3$ factors: $\CW_{10}(\blambda)$,
$\CW_8(\blambda) \times \RR^2$, $\CW_6(\blambda) \times \RR^4$,
$\CW_4(\blambda) \times \RR^6$ and $\RR^{1,9}$.

The gaugino variation says that
\begin{equation*}
 F \varepsilon = f_i dx^i \wedge dx^- \varepsilon = 0~,
\end{equation*}
which is equivalent, for nonzero $F$, to $dx^- \varepsilon = 0$; and
on such spinors, the dilatino equation is automatically satisfied.

In summary, Killing spinors are in one-to-one
correspondence with the subspace of the chiral spinor representation
of $\Spin(1,9)$ consisting of spinors which are annihilated by $dx^-$.
This is clearly an $8$-dimensional subspace, whence these backgrounds
are $\half$-BPS.

\begin{table}[h!]
  \centering
  \setlength{\extrarowheight}{3pt}
  \renewcommand{\arraystretch}{1.3}
  \begin{small}
    \begin{tabular}{|>{$}l<{$}|>{$}c<{$}|>{$}c<{$}|>{$}c<{$}|}\hline
      \multicolumn{1}{|c|}{Parallelisable} &
      \multicolumn{3}{c|}{Supersymmetries with dilaton being}\\
      \multicolumn{1}{|c|}{geometry} &
      \multicolumn{1}{c|}{constant} &
      \multicolumn{1}{c|}{linear} &
      \multicolumn{1}{c|}{nonlinear} \\
      \hline\hline
      \AdS_3 \times S^3 \times S^3 \times \RR &  8 & 8 & \times \\
      \AdS_3 \times S^3 \times \RR^4 & 8 & 8 & \times \\
      \RR^{1,1} \times \SU(3) & \times & 8 & \times \\
      \RR^{1,3} \times S^3 \times S^3 & \times & 8 & 4\\
      \RR^{1,6} \times S^3 & \times & 8 & 4\\
      \RR^{1,9} & 16 & 8 & 8\\
      \CW_{10}(\blambda) & 8, 10, 12, 14 & 8 & 8\\
      \CW_8(\blambda) \times \RR^2& 8, 10 & 8 & 8\\
      \CW_6(\blambda) \times S^3 \times \RR & \times & 8 & 4\\
      \CW_6(\blambda) \times \RR^4 & 8, 12 & 8 & 8 \\
      \CW_4(\blambda) \times S^3 \times \RR^3 & \times & 8 & 4\\
      \CW_4(\blambda) \times \RR^6 & 8 & 8 & 8\\
      \hline
    \end{tabular}
  \end{small}
  \vspace{8pt}
  \caption{Supersymmetric parallelisable backgrounds}
  \label{tab:summary}
\end{table}

\section{On the moduli space of parallelisable backgrounds}
\label{sec:moduli}

The backgrounds in Table~\ref{tab:summary} each come with moduli and
taken in their totality comprise the moduli space of simply-connected
parallelisable backgrounds.  This moduli space is infinite-dimensional
due to the arbitrary functions entering in the expressions for the
dilaton and the gauge fields, when nonzero.  There are also geometric
moduli: the radii of curvature of the $\AdS_3$, $S^3$ and $\SU(3)$
factors, and the eigenvalues $\blambda$ appearing in the definition of
$\CW_{2n}(\blambda)$.  Focussing on the geometric moduli for
simplicity, we remark that all the backgrounds are connected by the
following geometric limits:
\begin{itemize}
\item $\SU(3) \rightsquigarrow \RR^8$ by taking the radius of
  curvature to infinity;
\item $S^3 \rightsquigarrow \RR^3$ by taking the radius of curvature
  to infinity;
\item $\AdS_3 \rightsquigarrow \RR^{1,2}$ by taking the radius of
  curvature to infinity;
\item $\CW_{2n}(\lambda_1,\dots,\lambda_{n-1}) \rightsquigarrow
  \CW_{2n-2}(\lambda_1, \dots, \lambda_{n-2}) \times \RR^2$ by taking
  $\lambda_{n-1} \to 0$;
\item $\AdS_3 \times S^3 \times \RR^4 \rightsquigarrow \CW_6(\blambda)
  \times \RR^4$ by taking a plane wave limit
  \cite{PenrosePlaneWave,GuevenPlaneWave}; and
\item $\AdS_3 \times S^3 \times S^3 \times \RR \rightsquigarrow
  \CW_8(\blambda) \times \RR^2$ again by a plane wave limit.
\end{itemize}
Some of these plane wave limits have appeared in
\cite{ShortLimits,Limits}.  As the above geometries are parallelised
Lie groups, these plane wave limits can also be understood as group
contractions \cite{ORS,FSPL} in the sense of In\"on\"u and Wigner
\cite{InonuWigner}.

Finally we should remark that we have classified the simply-connected
backgrounds. Equivalently we do not distinguish between backgrounds
which are locally isometric; for example, two backgrounds which are
obtained as different discrete quotients of the same simply-connected
background.  More generally, in order to classify all smooth
backgrounds, one must quotient the simply-connected geometries in
Table~\ref{tab:summary} by all possible freely-acting discrete
subgroups of symmetries which preserve some supersymmetry and are free
of singularities.  This is a much more delicate problem and we will
only mention the fact that performing this quotient corresponds, at
the level of the conformal field theory, to an orbifold construction,
whence the string theory remains, in principle, exactly solvable.

\section*{Acknowledgments}

We are grateful to Shyamoli Chaudhuri, Yosuke Imamura, Tom\'as Ort\'in
and Yuji Tachikawa for helpful correspondence and discussions.  This
work was completed while JMF was visiting CERN, whom he would like to
thank for support and hospitality, and TK and SY were participating at
the Summer Institute ``Fuji 2003'' in Shizuoka, Japan, whose organisers they
would like to thank for hospitality.  The research of JMF is partially
funded by the UK EPSRC grant GR/R62694/01 and that of SY by a JSPS
Research Fellowship for Young Scientists.  JMF is also a member of
EDGE, Research Training Network HPRN-CT-2000-00101, supported by The
European Human Potential Programme.

\appendix

\section{Clifford algebra conventions}
\label{sec:clifford}

Our Clifford algebra conventions mostly follow the book \cite{LM}, but
we will review them here briefly.

Let $\RR^{t,s}$ denote the real $(t+s)$-dimensional vector space with
inner product obtained from the norm
\begin{equation*}
  |\bx|^2 = -(x^1)^2 - \cdots - (x^t)^2 + (x^{t+1})^2 + \cdots +
  (x^{t+s})^2~,
\end{equation*}
for $\bx=(x^1,\dots,x^{s+t}) \in \RR^{t,s}$.  By definition the real
Clifford algebra $\Cl(t,s)$ is generated by $\RR^{t,s}$ (and the
identity $\1$) subject to the Clifford relation
\begin{equation*}
  \bx \cdot \bx  = - |\bx|^2 \1~,
\end{equation*}
where we ask the reader to pay close attention to the sign!  Let
$\be_a \in \RR^{t,s}$ be the elementary basis vectors relative to
which $\bx = \sum x^a \be_a$ and let $\Gamma_a$ denote their image in 
$\Cl(t,s)$.  Then the Clifford relations become
\begin{equation*}
  \Gamma_a \Gamma_b + \Gamma_b \Gamma_a = - 2 \eta_{ab} \1~,
\end{equation*}
where $\eta_{ab} = \left< \be_a,\be_b \right>$ are the components of
the inner product relative to this basis.

We are interested in ten-dimensional lorentzian signature: $\RR^{1,9}$
with orthonormal basis $\be_0,\be_1,\cdots,\be_9$ with $|\be_0|^2 =-1$
and $|\be_i|^2 = 1$ for $i=1,\dots,9$.  As a real associative algebra,
$\Cl(1,9)$ is isomorphic to the algebra of $32 \times 32$ real
matrices.  This is a simple algebra and hence has a unique irreducible
representation $\eS$, which is real and thirty-two dimensional, the
so-called Majorana spinors.  The chirality operator $\Gamma_{11} :=
\Gamma_0 \Gamma_1 \cdots \Gamma_9$ squares to $+\1$ and anticommutes
with all $\Gamma_a$, whence it commutes with the generators $\half
\Gamma_{ab}$ of $\fso(1,9)$.  Therefore $\eS$ breaks up into two real
sixteen-dimensional irreducible representations of the spin group
$\Spin(1,9)$: $\eS = \eS_+ \oplus \eS_-$, the so-called Majorana--Weyl
spinors.

The Clifford algebra $\Cl(1,9)$ is isomorphic as a real vector space
(but \emph{not} as an algebra) to the exterior algebra
$\Lambda\RR^{1,9}$, whose elements are real linear combinations of
monomials of the form $\be_{a_1} \wedge \be_{a_2} \wedge \cdots \wedge
\be_{a_k}$.  There is a natural isomorphism $\Lambda\RR^{1,9} \to
\Cl(1,9)$ given by sending
\begin{equation}
  \label{eq:iso}
  \be_{a_1} \wedge \be_{a_2} \wedge \cdots \wedge \be_{a_k} \mapsto
  \Gamma_{a_1 a_2 \cdots a_k}~.
\end{equation}
This map also preserves the canonical $\ZZ_2$ gradings of $\Cl(1,9)$
and of $\Lambda\RR^{1,9} = \Lambda^{\text{even}}\RR^{1,9} \oplus
\Lambda^{\text{odd}}\RR^{1,9}$.  In this way, elements of
$\Lambda\RR^{1,9}$ can act on $\eS$: even elements preserving the
chirality, whence mapping $\eS_\pm \to \eS_\pm$, and odd elements
reversing it, whence mapping $\eS_\pm \to \eS_\mp$.

Now let $(M,g)$ be a lorentzian ten-dimensional manifold.  We can
choose local orthonormal frames for the tangent bundle $TM$ and dual
coframes for the cotangent bundle $T^*M$.  Relative to such a coframe,
each cotangent space is isomorphic to $\RR^{1,9}$ as an inner product
space and we can construct at each point a Clifford algebra
$\Cl(1,9)$.  As we let the point vary, these algebras patch up nicely
to yield a bundle $\Cl(T^*M)$ of Clifford algebras which, as a vector
bundle, is isomorphic to $\Lambda T^*M$.  The isomorphism
\eqref{eq:iso} also extends to give a map $\Lambda T^*M \to
\Cl(T^*M)$.

If in addition, $(M,g)$ is spin, then there is a (not necessarily
unique) vector bundle $S$ associated to the irreducible representation
$\eS$ of $\Cl(1,9)$.  Furthermore this bundle breaks up into
sub-bundles $S = S_+ \oplus S_-$ corresponding to the irreducible
representations of the spin group.  Sections of $\Lambda T^*M$---that
is, differential forms--- act naturally on sections of $S$ via the
isomorphism $\Lambda T^*M \to \Cl(T^*M)$ and the natural pointwise
action of $\Cl(T^*M)$ on $S$.  This action extends to an action of
forms with values in a vector bundle $V$ which maps sections of $S$ to
sections of $S \otimes V$.  In our case, $V$ will be the adjoint
bundle of the gauge bundle.  This now explains what we mean by the
action of forms on spinors, as in equation \eqref{eq:susyvars}.

Several times during the calculations in this paper we have come
across Clifford squares; that is, the repeated action of a
differential form on a spinor.  For $\alpha$ a $1$-form, we obtain
simply
\begin{equation*}
  \alpha^2 = - |\alpha|^2 \1~;
\end{equation*}
whereas for $\omega$ a $2$-form, we find
\begin{equation*}
  \omega^2 = \omega \wedge \omega - |\omega|^2 \1~.
\end{equation*}
For $H$ a $3$-form we find
\begin{equation*}
  H^2 = - \tfrac14 H_{abm}H^m{}_{cd} \Gamma^{abcd}
  + |H|^2 \1~,
\end{equation*}
which, if $H$ satisfies the Jacobi identity, simplifies to
\begin{equation*}
  H^2 = |H|^2 \1~.
\end{equation*}
Finally, if $\Theta$ is a $4$-form,
\begin{equation*}
  \Theta^2 = \Theta \wedge \Theta - \tfrac18 \Theta_{abmn}
  \Theta^{mn}{}_{cd} \Gamma^{abcd} + |\Theta|^2 \1~.
\end{equation*}

\providecommand{\href}[2]{#2}\begingroup\raggedright\endgroup

\end{document}